\documentclass[journal, twocolumn]{IEEEtran}

\usepackage{cite}
\usepackage{amsmath,amssymb,amsfonts}
\usepackage{gensymb}
\usepackage{float}
\usepackage{stfloats}
\usepackage{cancel}

\usepackage{array}
\usepackage{algorithmic}
\usepackage{graphicx}
\usepackage{textcomp}
\usepackage{xcolor}
\usepackage[hidelinks]{hyperref}
\usepackage{cancel}

\usepackage{amsthm}
\theoremstyle{definition}

\ifCLASSOPTIONcompsoc
    \usepackage[caption=false, font=normalsize, labelfont=sf, textfont=sf]{subfig}
\else
\usepackage[caption=false, font=footnotesize]{subfig}
\fi

\DeclareMathOperator{\sinc}{sinc}
\DeclareMathOperator{\rect}{rect}
\DeclareMathOperator{\sgn}{sgn}
\DeclareMathOperator{\Si}{Si}

\begin{document}

\title{Analysis and Approximation of a Spatially Wideband Antenna Array Factor}
 
\author{
    Marcin Wachowiak,~\IEEEmembership{Member,~IEEE,} 
    André Bourdoux,~\IEEEmembership{Senior~Member,~IEEE,}
    Sofie Pollin,~\IEEEmembership{Senior~Member,~IEEE,}%
    
    \thanks{
        Marcin Wachowiak and Sofie Pollin are with Interuniversitair Micro Electronica Centrum, 3001 Leuven, Belgium and also with the Katholieke Universiteit Leuven, 3000 Leuven, Belgium (e-mail: marcin.wachowiak@imec.be)}%
    \thanks{
        André Bourdoux is with Interuniversitair Micro-Electronica Centrum, 3001 Leuven, Belgium. (Corresponding author: \textit{Marcin Wachowiak})}
}

\maketitle

\begin{abstract}
This work investigates the spatially wideband (SWB) antenna array factor (AF) of uniform linear arrays. First, the SWB AF approximation is derived for narrowband (NB) signals and a large aperture with element spacing satisfying the Nyquist criterion. The derivation accounts for different spatial and spectral windows. Next, an approximation of the SWB AF for a wideband (WB) signal is developed under uniform spatial and spectral weighting. The analysis shows that for fully populated arrays, increasing the bandwidth effectively suppresses the AF sidelobes. Finally, a universal SWB AF approximation is introduced, which is based on recognizing the SWB AF as a spatially variant convolution. In this formulation, the SWB AF is expressed as a convolution of the spatially narrowband (SNB) AF and the SWB kernel, providing insight into how the bandwidth and the spectral weighting affect the resulting SWB AF. The proposed approximation is shown to be accurate for a wide range of bandwidths and element spacings, including sparse arrays. In particular, for sparse arrays, the bandwidth enables suppression of grating-lobe amplitudes by spreading their energy over a wider angular range. An approximation of the grating lobe envelope as a function of the bandwidth-aperture product is provided.
\end{abstract}


\section{Introduction}
\subsection{Problem Statement}
The array factor (AF) is an essential component of the signal model for systems with multiple antennas. It offers insight into the spatial properties of the system, such as the beam pattern and sidelobe levels \cite{van_trees_arrays, phased_arr_handbook, balanis_antenna}. Despite being a single term in the complex signal model and processing chain, it offers valuable high-level insight into the system expected performance and can drive the system-level design. 
For example, in a communication system, the AF determines the spatial selectivity, separation and interference between users \cite{massive_mimo_fundamentals}. 
In radar systems, the AF determines angular resolution and multiple-target performance based on the sidelobe levels \cite{principles_of_modern_radar}.

Typically, the antenna arrays are designed to operate over bandwidths that are inversely proportional to the array size \cite{phased_arr_handbook, balanis_antenna}.
To facilitate new applications such as joint communication and sensing, high-throughput communications and high-resolution radar imaging, the antenna arrays are expected to scale both in aperture size and in supported bandwidth \cite{mimo_distributed_radar, 6g_large_arr_tutorial, gigantic_mimo}. As the bandwidth and aperture of the antenna array increase, the spatially narrowband (SNB) condition is no longer satisfied and spatially wideband (SWB) effects must be considered and modeled accurately.
Note that in the context of antenna arrays the terms narrowband (NB) and wideband (WB) generally refer to the product of the spatial extent of the array and the signal bandwidth, rather than to the signal bandwidth alone. Here, to explicitly emphasize the spatial-bandwidth criterion, the terms SNB and SWB are used when referring to the AF.

\subsection{Relevant works}
The SWB effects on the communication system performance have been a subject of recent research with a focus on mitigating the beam squint and proposing WB precoders.
In \cite{spatial_wideband_mmw_mmimo, spatial_wideband_mmw_mmimo_app} the spatial-wideband effects are investigated in millimeter-wave massive multiple-input multiple-output (MIMO) systems from the wireless channel perspective. 
Wideband hybrid beamforming for WB MIMO terahertz communications systems is analyzed in \cite{wideband_hybrid_beamforming}.

On the other hand, the radar-centric research efforts have been centered around analysing the WB ambiguity function and proposing methods to suppress the grating lobes and sidelobes of the AF.
In \cite{nanzer_spatial_filtering_grating_lobes} the impact of bandwidth on spatial filtering of the grating lobes is investigated. The paper considers only two widely separated antennas in a mobile array scenario.
Sparse ultra-wideband (UWB) antenna arrays are discussed in \cite{bw_requirement_for_suppresing_grating_lobes}. The paper focuses on a small array with a low number of antenna elements and large relative bandwidths. 
In \cite{mimo_radar_amb_func} the ambiguity function of MIMO radar is extended to include the spatial aspect determined by the geometry of the sensors. The WB beampattern synthesis for MIMO radar by optimizing the spectral properties of the transmitted waveforms is discussed in \cite{beampattern_synth_for_wideband_mimo_radar}.
The ambiguity function of WB radars is analyzed in \cite{af_of_wideband_radars}.

The implications of a WB signal on the AF with a low number of antenna elements have been a topic of interest in UWB research.
In \cite{uwb_mimo_af_factoriability} the factorability of the UWB MIMO ambiguity function is discussed.
The performance of ultra-sparse and UWB antenna arrays is studied in \cite{ultra_sparse_ultra_wideband_arrays}.
A simplified approach to SWB AF can be found in frequency-diverse arrays research, where instead of continuous wide bandwidth, only a few widely spaced tones are considered.
The use of frequency diversity to suppress the grating lobes in the beampattern and ambiguity function is explored in \cite{fda_mimo_grating_lobe_suppression, fda_grating_lobe_suppression_sep_ap}.

Grating lobes and SWB effects are also crucial in the context of radar imaging, where they largely determine the imaging resolution and quality.
In \cite{gl_suppression_method_for_nf_distributed_arr}
different subarray configurations are explored for grating lobe suppression in imaging radar.
In \cite{grating_lobe_supp_for_mimo_imaging_arr} the shift of the grating lobe position in imaging is exploited to reduce their amplitude.
In \cite{gl_supp_in_nf_mimo_zero_migration} the broadening of the zeros of the array beampattern is used to reduce the grating and sidelobes in the near-field scenario.

Despite detailed analyses and signal models, the listed works do not provide a general closed-form expression for the SWB antenna AF. Such an expression is important for understanding how the signal bandwidth and the amplitude spectrum modify the SNB AF and lead to the resulting SWB AF.

\subsection{Contributions}
This work extends the concept of the AF to the SWB regime.
Throughout the work, several approximations of the SWB AF are proposed. First, the AF for large apertures with sampling satisfying the Nyquist criterion and NB signals is introduced. Next, the approximation AF of a WB signal under uniform spectral and spatial weighting is derived. Finally, a general approximation of the SWB AF is formulated for arbitrary element spacing and bandwidth. The SWB AF is rewritten as a spatially variant convolution of the SNB AF and a SWB kernel. This formulation offers an intuitive interpretation of how the bandwidth spectral windowing shapes the SWB AF. The proposed approximation is shown to be accurate for a wide range of parameters. In addition, the suppression of the grating lobes in sparse arrays through bandwidth is analyzed and an approximation of the grating lobe envelope as a function of the bandwidth-aperture product is provided.
The proposed framework provides several closed-form approximations of the SWB AF, allowing for more informed performance evaluation and system design.

\section{Antenna array factor}

\subsection{Spatially narrowband array factor (SNB AF)}
Consider an antenna array with $M$ antenna elements indexed by $m$, the set of indices is denoted by $\mathcal{M}$. 
The array is designed for center frequency $f_{\mathrm{c}}$ and the antennas are modeled as omnidirectional point radiators. Consider an arbitrary point of interest given by a vector $\mathbf{p}$. The distance from the $m$-th antenna to the point of interest is denoted by $d_m(\mathbf{p})$. 
Given the beamforming position denoted by $\mathbf{p}'$, the SNB AF can be written as
\begin{align}
    \label{eq:nb_af}
    \mathrm{AF_{SNB}}(\mathbf{p}', \mathbf{p}, f_{\mathrm{c}}) 
    &= \sum_{m \in \mathcal{M}} 
    w(m) e^{-j2\pi \frac{f_{\mathrm{c}}}{c} d_m(\mathbf{p}') } e^{j2\pi \frac{f_{\mathrm{c}}}{c} d_m( \mathbf{p}) } \nonumber \\
    &= \sum_{m \in \mathcal{M}} w(m)
    e^{-j2\pi \frac{f_{\mathrm{c}}}{c} \Delta d_m(\mathbf{p}', \mathbf{p}) }.
\end{align}
where $w(m)$ is the amplitude weight per antenna element and $\Delta d_m(\mathbf{p}', \mathbf{p}) = d_m(\mathbf{p}') - d_m(\mathbf{p})$ is the distance difference between $\mathbf{p}'$ and $\mathbf{p}$ for $m$-th antenna.
Note that the SNB AF can be viewed as a weighted NB matched filter.
The expression in \eqref{eq:nb_af} is valid for a single frequency, but can be considered accurate when the maximum propagation delay across the aperture is much smaller than the coherence time $\tau_D \ll \tau_B$. For an array aperture $D$, the maximum delay introduced by the antenna array is $\tau_D = D / c$, where $c$ is the speed of light. For a signal of bandwidth $B$, the coherence time is given by $\tau_B = 1/B$.
The SNB AF can serve as a good approximation as long as the SNB criterion is satisfied $\frac{D B}{c} \ll 1$. 
Defining the fractional bandwidth as $B_{\mathrm{f}} = B / f_{\mathrm{c}}$ and the normalized aperture as $D_{\lambda} = D  / \lambda_{\mathrm{c}}$ the SNB array condition can be rewritten as $D_{\lambda}B_{\mathrm{f}} \ll 1$, where $\lambda_{\mathrm{c}} = c / f_{\mathrm{c}}$. For example, for a fractional bandwidth of 10\% the aperture should be much smaller than $10 \lambda_{\mathrm{c}}$ and for a larger aperture $D=100\lambda_{\mathrm{c}}$, the fractional bandwidth $B_{\mathrm{f}}$ should be much smaller than 1\%.
As the aperture size or operating bandwidth increases, the SNB AF becomes inaccurate, and the SWB AF must be considered.

\subsection{Spatially wideband array factor (SWB AF)}
Consider that each antenna transmits or receives a signal of bandwidth $B$ at the center frequency $f_{\mathrm{c}}$. The signal is defined in the frequency domain as $S(f)$ with the total energy normalized to unity $\int_{-B/2}^{B/2} |S(f)|^2 \, df = 1$.
The SWB AF is defined following a matched filter formulation. To account for the signal bandwidth in the SWB AF, the matching must be performed jointly in the spatial and frequency domains. The signal $S(f)$ is matched with its complex conjugate, yielding $|S(f)|^2$. The SWB AF is obtained by integrating the SNB AF \eqref{eq:nb_af} over the signal bandwidth, weighted by the signal energy spectral density $|S(f)|^2$
\begin{align}
    \label{eq:wb_af}
    &\mathrm{AF_{SWB}}(\mathbf{p}', \mathbf{p}) 
    = \int_{-B/2}^{B/2} S(f)S(f)^* 
    \mathrm{AF_{SNB}}(\mathbf{p}', \mathbf{p}, f_{\mathrm{c}} + f) \, df \nonumber  \\
    & \quad= \int_{-B/2}^{B/2} \left| S(f)\right|^2
    \sum_{m \in \mathcal{M}}  w(m)e^{-j2\pi \frac{f_{\mathrm{c}} + f}{c} \Delta d_m(\mathbf{p}', \mathbf{p}) } \, df.
\end{align}
Separating the carrier frequency and the frequency variable components gives
\begin{align}
    \label{eq:wb_af_sep}
    \mathrm{AF_{SWB}}(\mathbf{p}', \mathbf{p}) 
    &= \sum_{m \in \mathcal{M}} w(m) e^{-j2\pi \frac{f_{\mathrm{c}}}{c} \Delta d_m(\mathbf{p}', \mathbf{p}) } \\
    & \quad \times \int_{-B/2}^{B/2} |S(f)|^2
    e^{-j2\pi \frac{f}{c} \Delta d_m(\mathbf{p}', \mathbf{p}) } \, df. \nonumber
\end{align}

This work focuses on uniform linear antenna array geometries as they represent the most fundamental and widely used configuration. 
In the far-field of a linear array, the point of interest $\mathbf{p}$ simplifies to the distance to the array weight center and angle with respect to the array normal $\mathbf{p} = [d, \theta]$.
The distance from $m$-th antenna to point $\mathbf{p}$ is
\begin{align}
    \label{eq:dist_per_ant}
    d_m(\mathbf{p})
    &= d - md_{\mathrm{s}} \sin{(\theta)}, 
\end{align}
where $d_{\mathrm{s}}$ is the antenna element spacing.
Matching to the transmitted or received signal in \eqref{eq:wb_af}, introduces range dependency (resolution) in the SWB AF due to the waveform bandwidth providing temporal resolution.
To keep the conventional far-field interpretation of the AF and focus on the angular cross-cut of the AF, $\mathbf{p}$ and $\mathbf{p}'$ are fixed to the same distance $d'$.
The distance difference per antenna element, given that $d = d'$, is
\begin{align}
    \label{eq:dist_diff_per_ant}
    \Delta d_m(\Delta u) 
    & = -m d_{\mathrm{s}} \Delta u,
\end{align}
where $\Delta u = u' - u = \sin{(\theta')} - \sin{(\theta)} $ is the difference in the  direction sine ($u$) domain.
After substituting \eqref{eq:dist_diff_per_ant} into \eqref{eq:wb_af_sep}, the SWB antenna AF for the uniform linear antenna array is
\begin{align}
    \label{eq:wb_af_ula}
    \mathrm{AF_{SWB}}(\Delta u) 
    &= \sum_{m \in \mathcal{M}} w(m) e^{j2\pi m \frac{d_{\mathrm{s}}}{\lambda_{\mathrm{c}}} \Delta u} \\
    & \quad \times \int_{-B/2}^{B/2} |S(f)|^2
    e^{j2\pi m \frac{f}{c} d_{\mathrm{s}} \Delta u} \, df. \nonumber
\end{align}
The general expression of the ULA SWB AF derived in \eqref{eq:wb_af_ula} does not offer straightforward insight regarding the impact of wider bandwidth or larger aperture on the resultant SWB AF. 
The following subsections provide approximations of the SWB AF for different cases, offering actionable insight into sidelobe levels of the SWB AF.

\section{Approximations of spatially wideband  array factor for $d_{\mathrm{s}} \leq \lambda_{\mathrm{c}}/2$}

\subsection{Spatially wideband array factor with narrowband signal}

Consider a uniform linear antenna array (ULA) with element spacing satisfying the Nyquist criterion $d_{\mathrm{s}} \leq \lambda_{\mathrm{c}}/2$.
Given that the antennas are distributed symmetrically around the weight center of the array, the sum over antenna elements in \eqref{eq:wb_af} becomes
\begin{align}
    \label{eq:la_nb_full}
    & \mathrm{AF_{SWB}}(\Delta u) = 
    \int_{-B/2}^{B/2} |S(f)|^2  \nonumber \\
    & \qquad \times \sum_{m=-(M-1)/2}^{(M-1)/2} w(m) e^{j 2\pi m \frac{f_{\mathrm{c}} + f}{c}  d_{\mathrm{s}} \Delta u }\, df.
\end{align}
The spatial window $w(m)$ can be expressed in terms of the discrete Fourier transform (DFT) coefficients as 
\begin{align}
    \label{eq:wdw_fft}
    w(m) = \frac{1}{M} \sum_{n=-(M-1)/2}^{(M-1)/2} W(n) e^{j2\pi m \frac{n}{M}},
\end{align}
where $n$ denotes the spatial frequency index.
Isolating the sum corresponding to the AF in \eqref{eq:la_nb_full} and expanding it with \eqref{eq:wdw_fft} gives
\begin{align}
    \label{eq:wdw_exp_expanded}
    \mathrm{AF_{SNB}}(\Delta u, f_{\mathrm{c}} +f) 
    &= \frac{1}{M}\sum_{n=-(M-1)/2}^{(M-1)/2} 
    W(n) \\
    & \quad \times \sum_{m=-(M-1)/2}^{(M-1)/2}e^{j 2\pi m \left(\frac{f_{\mathrm{c}} + f}{c} d_{\mathrm{s}} \Delta u + \frac{n}{M} \right) }, \nonumber 
\end{align}
where $W(n)$ is the DFT of the spatial window.
The symmetric sum over antenna indices in \eqref{eq:wdw_exp_expanded} can be expressed in terms of the Dirichlet kernel as follows
\begin{align}
    \label{eq:dirichlet_kernel}
    &\sum_{m=-(M-1)/2}^{(M-1)/2}e^{j 2\pi m \left(\frac{f_{\mathrm{c}} + f}{c} d_{\mathrm{s}} \Delta u + \frac{n}{M} \right) } = \nonumber \\
    &\qquad = \frac{\sin{\left( \pi M  \left( d_{\mathrm{s}} \frac{f_{\mathrm{c}} + f}{c}\Delta u + \frac{n}{M} \right) \right)}}
    {\sin{\left( \pi \left( \frac{d_{\mathrm{s}}}{\lambda_{\mathrm{c}}}\left(1 + \frac{f}{f_{\mathrm{c}}}\right) \Delta u+ \frac{n}{M} \right) \right)}}. 
\end{align}
Expressing the window in terms of DFT \eqref{eq:wdw_exp_expanded} and applying the Dirichlet kernel identity \eqref{eq:dirichlet_kernel} to \eqref{eq:la_nb_full} yields
\begin{align}
    \label{eq:la_nb_dirichlet}
    & \mathrm{AF_{SWB}}(\Delta u) = 
    \quad = \frac{1}{M} \sum_{n=-(M-1)/2}^{(M-1)/2} 
    W(n) \\
    &\quad \times \int_{-B/2}^{B/2} |S(f)|^2  
    \frac{\sin{\left( \pi M  \left( d_{\mathrm{s}} \frac{f_{\mathrm{c}} + f}{c}\Delta u + \frac{n}{M} \right) \right)}}
    {\sin{\left( \pi \left( \frac{d_{\mathrm{s}}}{\lambda_{\mathrm{c}}}\left(1 + \frac{f}{f_{\mathrm{c}}}\right) \Delta u+ \frac{n}{M} \right) \right)}}\, df \nonumber.
\end{align}
To proceed, assume that signal $S(f)$ is narrowband $B / f_{\mathrm{c}} \ll 0.1$ and $M$ antennas implement a large aperture given by $D = (M-1) d_{\mathrm{s}}$. Under this assumption, the $f / f_{\mathrm{c}}$ component in the denominator in \eqref{eq:la_nb_dirichlet} can be considered negligible, leading to
\begin{align}
    \label{eq:la_nb_simp}
    & \mathrm{AF_{SWB}}(\Delta u) 
    \overset{B\ll f_{\mathrm{c}}}{\approx} 
    \frac{1}{M} \sum_{n=-(M-1)/2}^{(M-1)/2} 
    \frac{W(n)}{\sin{\left( \pi \left(\frac{d_{\mathrm{s}}}{\lambda_{\mathrm{c}}} \Delta u + \frac{n}{M}\right) \right)}} \nonumber \\
    & \times \int_{-B/2}^{B/2} |S(f)|^2  
    \sin{\left( \pi M \left(\frac {d_{\mathrm{s}}}{\lambda_{\mathrm{c}}} \Delta u + \frac{f}{c}  d_{\mathrm{s}} \Delta u  + \frac{n}{M}\right) \right)}
     \, df. 
\end{align}
Next, by applying the sum-to-product trigonometric identity in \eqref{eq:la_nb_simp} and assuming that $|S(f)|^2$ is even, the integral of the sine term over symmetric bounds vanishes, yielding
\begin{align}
    \label{eq:la_nb_cos_int}
    & \mathrm{AF_{SWB}}(\Delta u) = 
    \int_{-B/2}^{B/2} |S(f)|^2 
    \cos{\left( \pi f \frac{M d_{\mathrm{s}} \Delta u }{c} \right)}
    \, df \\
    & \qquad \times  
    \frac{1}{M} \sum_{n=-(M-1)/2}^{(M-1)/2} W(n)
    \frac{\sin{\left( \pi M \left( \frac{d_{\mathrm{s}}}{\lambda_{\mathrm{c}}}\Delta u + \frac{n}{M} \right) \right)} }
    {\sin{\left( \pi \left(\frac{d_{\mathrm{s}}}{\lambda_{\mathrm{c}}} \Delta u + \frac{n}{M}\right) \right)}}. \nonumber
\end{align}
For a deterministic signal, the autocorrelation function is given by the inverse Fourier transform of its energy spectral density $|S(f)|^2$ as follows
\begin{align}
    \label{eq:autocorr}
    R(\tau) = \int_{-B/2}^{B/2} |S(f)|^2 e^{j2\pi f \tau} \, df.
\end{align}
Based on the previous assumption that $|S(f)|^2$ is even the integral in \eqref{eq:la_nb_cos_int} can be expressed in terms of the autocorrelation function \eqref{eq:autocorr} as
\begin{align}
    \label{eq:sf_int}
   \int_{-B/2}^{B/2} |S(f)|^2 
    \cos{\left( \pi f \frac{M d_{\mathrm{s}} \Delta u }{c} \right)}
    \, df  &= R \left( \frac{M d_{\mathrm{s}} \Delta u }{2c} \right).
\end{align}
Note that the sum term in \eqref{eq:la_nb_cos_int} is equal to the SNB AF \eqref{eq:wdw_exp_expanded} at $f_{\mathrm{c}}$ expressed in terms of the Dirichlet kernel from \eqref{eq:dirichlet_kernel}.
This allows to write the SWB AF \eqref{eq:la_nb_cos_int} as a product of the waveform autocorrelation function \eqref{eq:sf_int} and the SNB AF \eqref{eq:wdw_exp_expanded} as follows
\begin{align}
    \label{eq:wb_af_nb_sig}
    \mathrm{AF_{SWB}}(\Delta u) 
    =& R \left( \frac{M d_{\mathrm{s}} \Delta u }{2c}  \right) \mathrm{AF_{SNB}}(\Delta u, f_{\mathrm{c}}).
\end{align}
This formulation captures both the array tapering and the weighting in the spectral domain and accurately approximates the SWB AF for systems with narrowband signals. Spatial windowing affects the total SWB AF by shaping the SNB AF, while the spectral windowing modifies the autocorrelation properties of the signal, affecting the SWB AF sidelobe decay.

\subsubsection{Uniform spatial weighting}
\begin{figure}[tb]
    \centering
    \includegraphics[width=\linewidth]{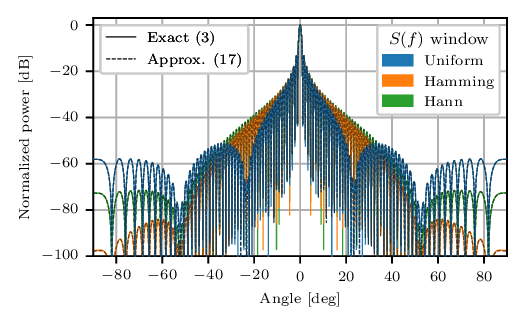}
    \caption{Normalized ULA AF and its approximation for uniform spatial weighting and different amplitude spectrum windows, $D=50 \lambda_{\mathrm{c}}$, $B_{\mathrm{f}} = 0.1$, $d_{\mathrm{s}} = \lambda_{\mathrm{c}}/2$. }
    \label{fig:af_uni_arr_wdw_sig}
\end{figure}

For uniform weighting across the antenna elements  $w(m) = \frac{1}{\sqrt{M}} $, constraining the total radiated energy to unity $\sum_m |w(m)|^2 = 1$, the SWB AF from \eqref{eq:wb_af_nb_sig} simplifies to
\begin{align}
    \label{eq:uni_taper_la_nb_af}
    \mathrm{AF_{SWB}}(\Delta u) 
    = & 
    R{\left( \frac{M d_{\mathrm{s}} \Delta u }{2c} \right)}
    \frac{1}{\sqrt{M}} \frac{\sin{\left( \pi M \frac{d_{\mathrm{s}}}{\lambda_{\mathrm{c}}} \Delta u \right)} }
    {\sin{\left( \pi \frac{d_{\mathrm{s}}}{\lambda_{\mathrm{c}}} \Delta u \right)}}.
\end{align}
Fig. \ref{fig:af_uni_arr_wdw_sig} illustrates the effect of different spectral windows on the SWB AF. The windowing in the spectral domain does not affect the width of the mainlobe in the angular domain; it only affects the shape and rate of decay of the sidelobes.

\subsubsection{Uniform amplitude spectrum}
\begin{figure}[tb]
    \centering
    \includegraphics[width=\linewidth]{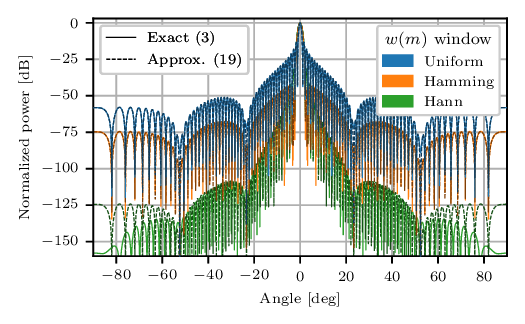}
    \caption{Normalized ULA AF and its approximation for uniform amplitude spectrum and different spatial windows, $D=50 \lambda_{\mathrm{c}}$, $B_{\mathrm{f}} = 0.1$, $d_{\mathrm{s}} = \lambda_{\mathrm{c}}/2$.}
    \label{fig:af_uni_sig_wdw_arr}
\end{figure}

For flat, uniform amplitude spectrum $|S(f)| = 1 / \sqrt{B}$ the integral in \eqref{eq:sf_int} yields a closed form-expression
\begin{align}
    \label{eq:flat_sf}
    R \left( \frac{M d_{\mathrm{s}} \Delta u }{2c} \right) 
    &= \sinc{\left( B \frac{M d_{\mathrm{s}} \Delta u }{2c} \right)},
\end{align}
where $\sinc{(x)} = \sin{(\pi x)} / (\pi x)$
and the SWB AF from \eqref{eq:wb_af_nb_sig} becomes
\begin{align}
    \label{eq:uni_sig_la_nb_af}
     \mathrm{AF_{SWB}}(\Delta u) 
    = & \sinc{\left( B \frac{M d_{\mathrm{s}} \Delta u }{2c} \right)} \mathrm{AF_{SNB}}(\Delta u, f_{\mathrm{c}}) . 
\end{align}
Fig. \ref{fig:af_uni_sig_wdw_arr} illustrates the effect of different spatial windows on the SWB AF. As expected, windowing in the spatial domain modifies the SNB AF by widening the mainlobe and suppressing sidelobe levels.

\subsubsection{Uniform spatial weighting and uniform amplitude spectrum}
\begin{figure}[tb]
    \centering
    \includegraphics[width=\linewidth]{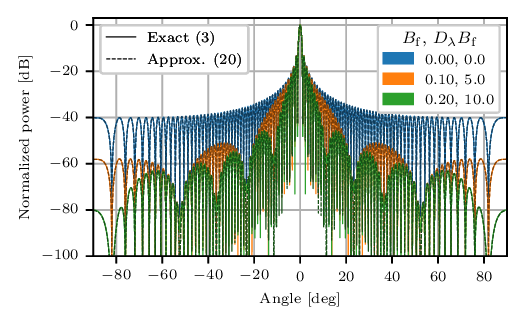}
    \caption{Normalized ULA AF and its approximation for uniform spatial weighting and uniform amplitude spectrum for different values of the fractional bandwidth $B_{\mathrm{f}}$, $D=50 \lambda_{\mathrm{c}}$, $d_{\mathrm{s}} =  \lambda_{\mathrm{c}}/2$.}
    \label{fig:af_uni_sig_uni_arr_vs_fbw}
\end{figure}

Under uniform array weighting and uniform amplitude spectrum conditions, the SWB AF from \eqref{eq:wb_af_nb_sig} simplifies to
\begin{align}
    \label{eq:uni_sig_uni_arr_nb_af}
     \mathrm{AF_{SWB}}(\Delta u) 
    = & \sinc{\left( B \frac{M d_{\mathrm{s}} \Delta u }{2c} \right)}  \frac{1}{\sqrt{M}} \frac{\sin{\left( \pi M \frac{d_{\mathrm{s}}}{\lambda_{\mathrm{c}}} \Delta u \right)} }
    {\sin{\left( \pi \frac{d_{\mathrm{s}}}{\lambda_{\mathrm{c}}} \Delta u \right)}}.
\end{align}
Fig. \ref{fig:af_uni_sig_uni_arr_vs_fbw} shows the impact of increasing the bandwidth-aperture product $D_{\lambda}B_{\mathrm{f}}$ on the total AF. For a half-wavelength-spaced array, the increase in the bandwidth-aperture product reduces the sidelobe levels.
The proposed approximation remains accurate with regard to the exact expression across different values of the bandwidth-aperture products and different windowing scenarios.

\subsection{Spatially wideband array factor for wideband signal, under uniform spatial and spectral weighting}

\begin{figure}[tb]
    \centering
    \includegraphics[width=\linewidth]{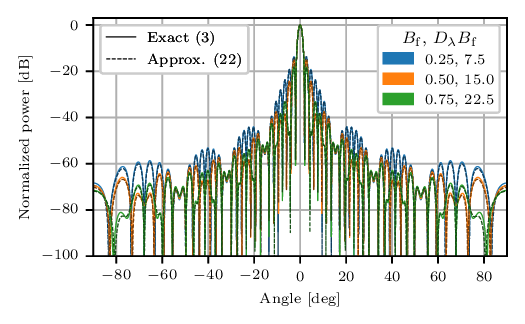}
    \caption{Normalized ULA AF and its approximation for uniform amplitude spectrum and uniform spatial weighting for different values of fractional bandwidth $B_{\mathrm{f}}$, $D = 30 \lambda_{\mathrm{c}}$, $d_{\mathrm{s}} = \lambda_{\mathrm{c}}/4$.}
    \label{fig:af_uni_wb_sig_uni_arr_vs_fbw}
\end{figure}

For fully populated and dense arrays, a closed-form expression for the SWB AF can be obtained by approximating the aperture as continuous.
However, arriving at this closed-form approximation requires constraining both the spatial weighting and the amplitude spectrum to be uniform as follows $w(m) = \frac{1}{\sqrt{M}} $ and $|S(f)| = \frac{1}{\sqrt{B}}$
For uniform array weighting and a uniform amplitude spectrum the integral over the bandwidth in \eqref{eq:wb_af_sep} reduces to a $\sinc$ function as follows
\begin{align}
    \label{eq:wb_af_full}
    \mathrm{AF_{SWB}}(\Delta u) 
    & = \frac{1}{\sqrt{M}} \sum_{-(M-1)/2}^{(M-1)/2} e^{j 2\pi  m \frac{d_{\mathrm{s}}}{\lambda_{\mathrm{c}}}  \Delta u } \sinc\left( \frac{B}{c} m d_{\mathrm{s}} \Delta u \right) .
\end{align}
The discrete sum over the antenna elements is approximated by an integral using the Riemann sum definition of the integral. This approximation is valid under the condition that the argument of the sum is smooth and slowly varying across $m$. This requires the change of $\sinc$ to be negligible as $m$ varies, yielding a following constraint $B_{\mathrm{f}} \ll 1 / d_{\mathrm{s}_{\lambda}}$. Assuming the number of antenna elements $M$ is large, so that the inter-element spacing becomes small, the discrete sum can approximate the integral well.
Moreover, the antenna spacing must satisfy the Nyquist criterion $(d_{\mathrm{s}} \leq \lambda_{\mathrm{c}}/2)$ to prevent aliasing and the introduction of grating lobes.
By applying the continuous aperture approximation in \eqref{eq:wb_af_full} and then exploiting the symmetry of the integration bounds, the complex exponential term can be recast as a cosine. Subsequently, by expressing the resulting trigonometric product as a sum and evaluating the integral, the SWB AF of \eqref{eq:wb_af_full} can be written as
\begin{align}
    \label{eq:wb_int_approx}
    &\mathrm{AF_{SWB}}(\Delta u) \overset{B_{\mathrm{f}} \ll 1 / d_{\mathrm{s}_{\lambda}}}{\approx}  \\
    &\overset{B_{\mathrm{f}} \ll 1 / d_{\mathrm{s}_{\lambda}}}{\approx}
    \frac{1}{\sqrt{M}} \int_{-M/2}^{M/2} e^{j 2\pi m \frac{d_{\mathrm{s}}}{\lambda_{\mathrm{c}}} \Delta u} \sinc\left( \frac{B}{c} m d_{\mathrm{s}} \Delta u \right) \, dm \nonumber \\
    &\qquad = \frac{1}{\sqrt{M}} \frac{1}{d_{\mathrm{s}}} \frac{1}{ \frac{B}{c} \Delta u }
    \Bigg( \Si{\left( \Delta u  (D + d_{\mathrm{s}})  \left( \frac{B}{2c} + 1 \right) \right)} \nonumber \\
    &\quad\qquad\qquad\qquad\qquad + \Si{\left( \Delta u  (D + d_{\mathrm{s}})  \left( \frac{B}{2c} - 1 \right) \right)} 
    \Bigg), \nonumber 
\end{align}
where $\Si{(x) = \int_{0}^{x} \sinc{(x)} \, dx} = \int_{0}^{x} \sin{(\pi x)} / (\pi x) \, dx$ is the normalized sine integral function. The approximation can be used for large bandwidths, given that the antenna spacing is sufficiently small.
Fig. \ref{fig:af_uni_wb_sig_uni_arr_vs_fbw} illustrates the SWB AF and its approximation for a WB signal and a quarter-wavelength spaced array. The approximation remains accurate over most of the angular region, with noticeable deviations near the edges corresponding to endfire angles. At large angles, the approximation fails to accurately reconstruct the sidelobe levels, as it is based on a continuous aperture approximation and does not capture well the periodicity of the AF due to the discrete nature of the antenna array.

\section{Universal approximation of spatially wideband array factor}
\label{sec:gen_wb_af_approx}
An alternative formulation of the SWB AF can be derived by recognizing it as a spatially variant convolution. Substituting the signal autocorrelation function from \eqref{eq:autocorr} into the decoupled SWB expression from \eqref{eq:wb_af_sep} gives
\begin{align}
    \label{eq:wb_af_r_sum}
    \mathrm{AF_{SWB}}(\Delta u) 
    &= \sum_{m \in \mathcal{M}} w(m) R \left( \frac{m d_{\mathrm{s}}\Delta u }{c} \right) e^{j 2\pi m \frac{d_{\mathrm{s}}}{\lambda_{\mathrm{c}}} \Delta u }.
\end{align}
In this form, the autocorrelation term $R \left( \frac{m d_{\mathrm{s}}\Delta u }{c} \right)$ acts as a spatially variant weighting across the array elements.
The product in \eqref{eq:wb_af_r_sum} can be expressed as a convolution using the discrete-time Fourier transform (DTFT). Expressing the SWB AF as a convolution enables its decomposition into the SNB AF and the SWB kernel, offering insight into how SWB effects modify the SNB AF to yield the SWB AF. The DTFT is defined as follows
\begin{align}
    \label{eq:dtft_def}
    \mathrm{DTFT} \left\{ a(m)\right\} (v) &= \sum_{m \in \mathcal{M}} a(m) e^{-j 2 \pi m v}.
\end{align}
The DTFT maps the discrete spatial domain denoted by $m$ to a continuous periodic function of the spatial frequency $v$.
To match the original expression from \eqref{eq:wb_af_r_sum}, with the DTFT definition in \eqref{eq:dtft_def}, the SWB AF is obtained by evaluating the DTFT for $v=-\frac{d_{\mathrm{s}}}{\lambda_{\mathrm{c}}} \Delta u$
\begin{align}
    \label{eq:prod_dtft}
    \mathrm{AF_{SWB}}(\Delta u)  = \mathrm{DTFT} \left\{w(m) R \left( \frac{m d_{\mathrm{s}}\Delta u }{c} \right) \right\} \left( -\frac{d_{\mathrm{s}}}{\lambda_{\mathrm{c}}} \Delta u \right).
\end{align}
The negative sign arises from an exponent sign mismatch between the original expression \eqref{eq:wb_af_r_sum} and the DTFT definition from \eqref{eq:dtft_def}.
The convolution theorem states that the DTFT of a product equals the periodic convolution of their individual DTFTs as follows
\begin{align}
    \label{eq:wb_af_dtft_conv}
    \mathrm{AF_{SWB}}(\Delta u)  
    = \Bigg( &\mathrm{DTFT} \left\{w(m)\right\} \\
    &\circledast
    \mathrm{DTFT}\left\{R \left( \frac{m d_{\mathrm{s}}\Delta u }{c} \right) \right\}
    \Bigg)
    \left( -\frac{d_{\mathrm{s}}}{\lambda_{\mathrm{c}}} \Delta u \right), \nonumber 
\end{align}
where $\circledast$ denotes periodic convolution.
The first factor in \eqref{eq:wb_af_dtft_conv} follows a formulation similar to that of a SNB AF
\begin{align}
    \label{eq:dtft_wm}
    \mathrm{DTFT} \left\{w(m)\right\} \left( v \right)  &= \sum_{m \in \mathcal{M}} w(m) e^{-j 2\pi m v} = W(v).
\end{align}
Evaluated for  $v = -\frac{d_{\mathrm{s}}}{\lambda_{\mathrm{c}}} \Delta u$ it becomes the SNB AF from \eqref{eq:nb_af} as follows $W\left( -\frac{d_{\mathrm{s}}}{\lambda_{\mathrm{c}}} \Delta u \right) = \mathrm{AF_{SNB}} ( \Delta u).$
The second factor in \eqref{eq:wb_af_dtft_conv} is a SWB kernel defined as
\begin{align}
    \label{eq:dtft_wb_kern}
    \mathrm{DTFT} \left\{ R \left( \frac{m d_{\mathrm{s}}\Delta u }{c} \right) \right\}  \left(v\right)
    &= \sum_{m \in \mathcal{M}} R \left( \frac{m d_{\mathrm{s}}\Delta u }{c} \right) e^{-j 2\pi m v} \nonumber \\
    &= R_{M}\left( \Delta u, v \right).
\end{align}
$R_{M}$ is a periodic continuous function of the spatial frequency $v$. 
Note that the first argument of $R_{M}$, which is $\Delta u$, is a fixed parameter, while $v$ is the convolution variable.
Given the DTFT of the convolution components \eqref{eq:dtft_wm} and \eqref{eq:dtft_wb_kern} the convolution in \eqref{eq:wb_af_dtft_conv} can be rewritten as
\begin{align}
    \label{eq:dtft_conv_int}
    \mathrm{AF_{SWB}}(\Delta u) &= \left( W \circledast
    R_{M}\left( \Delta u,  \cdot \right) \right)
    \left( -\frac{d_{\mathrm{s}}}{\lambda_{\mathrm{c}}} \Delta u \right) \\
    &= \int_{-1/2}^{1/2} 
    W\left( v \right)
    R_{M}\left( \Delta u, -\frac{d_{\mathrm{s}}}{\lambda_{\mathrm{c}}} \Delta u - v \right)\, dv \nonumber,
\end{align}
where $(\cdot)$ in $R_{M}$ is used to denote the convolution variable.
The periodic convolution has a period of 1 in the normalized spatial frequency $v$. The integration limits are chosen symmetrically around $0$. The SWB kernel $R_{\mathcal{M}}$ is spatially variant: it depends on $\Delta u$ through both the first and second arguments. Therefore, the resulting convolution is not shift invariant in $\Delta u$.
Applying a change of variable to express the convolution in the directional sine domain $u_{\tau}$ via substitution $v = - \frac{d_{\mathrm{s}} }{\lambda_{\mathrm{c}}}u_{\tau}$ and recognizing that $W\left( -\frac{d_{\mathrm{s}}}{\lambda_{\mathrm{c}}} \Delta u_{\tau} \right) = \mathrm{AF_{SNB}} (u_{\tau})$ allows to write the convolution as
\begin{align}
    \label{eq:dtft_conv_int_lim}
    \mathrm{AF_{SWB}}(\Delta u)
    &= \int_{-\lambda_{\mathrm{c}}/(2 d_{\mathrm{s}})}^{\lambda_{\mathrm{c}}/(2 d_{\mathrm{s}})} 
    \mathrm{AF_{SNB}}(u_{\tau}) \\
    & \qquad \times \frac{d_{\mathrm{s}}}{\lambda_{\mathrm{c}}} R_{M}\left( \Delta u, -\frac{d_{\mathrm{s}}}{\lambda_{\mathrm{c}}} \left( \Delta u - u_{\tau} \right) \right)\, du_{\tau}. \nonumber 
\end{align}

\subsection{Continuous approximation of spatially wideband kernel}

Calculating $R_{M}$ requires computing a discrete sum, which does not reduce to an insightful closed-form expression. 
To proceed, the function $R \left( \frac{m d_{\mathrm{s}}\Delta u }{c} \right)$ from \eqref{eq:dtft_wb_kern} is approximated by an integral using the Riemann sum definition of an integral. The discrete sum over antenna elements is replaced by a continuous integral over the aperture. This approximation can be considered accurate when the function $R \left( \frac{m d_{\mathrm{s}}\Delta u }{c} \right)$ varies slowly across the adjacent antenna positions. That is, the inter-element delay given by $\frac{d_{\mathrm{s}}\Delta u }{c}$ is small relative to the coherence time given by $1/B$, resulting in condition that can be written as $\frac{d_{\mathrm{s}} |\Delta u| }{c} \ll \frac{1}{B}$. Observing that the maximum absolute value of $|\Delta u|$ is $2$ and normalizing the antenna spacing to the wavelength $d_{\mathrm{s}_{\lambda}}= d_{\mathrm{s}} / \lambda_{\mathrm{c}}$ and the bandwidth in terms of fractional bandwidth $B_{\mathrm{f}}$, the approximation condition can be expressed as $B_{\mathrm{f}} \ll 1 / (2 d_{\mathrm{s}_{\lambda}})$.
Given that the indices of $M$ antenna elements are symmetrically distributed around 0 with index spacing $\Delta m = 1$, the integral approximation of a sum can be written as
\begin{align}
    \label{eq:rd_int_approx}
     &R_{M}\left( \Delta u, v \right)
     =\sum_{m \in \mathcal{M}} R \left( \frac{m d_{\mathrm{s}}\Delta u }{c} \right) e^{-j 2\pi m v} \Delta m\\
     &\qquad  \overset{B_{\mathrm{f}} \ll 1 / (2 d_{\mathrm{s}_{\lambda}}) }{\approx} \int_{-M/2}^{M/2}  R \left( \frac{ m d_{\mathrm{s}}\Delta u }{c} \right) e^{-j 2\pi m v} \, dm \nonumber.
\end{align}
The integral from \eqref{eq:rd_int_approx} can be expressed as an integral over the array aperture $D$ by a change of variable via substitution $x = m d_{\mathrm{s}}$, the new integration limits are $\pm M/2 = \pm(D + d_{\mathrm{s}}) / 2$
\begin{align}
    \label{eq:rd_ap_int}
    R_{(D + d_{\mathrm{s}})}\left( \Delta u, v \right)
    &= \frac{1}{d_{\mathrm{s}}} \int_{-(D + d_{\mathrm{s}}) / 2}^{(D + d_{\mathrm{s}}) / 2}  R \left( \frac{ x \Delta u }{c} \right) e^{-j 2\pi \frac{x}{d_{\mathrm{s}}} v} \, dx. 
\end{align}
The subscript $(D + d_{\mathrm{s}})$ in $R_D$ is used to denote the continuous approximation of the SWB kernel.

Approximating the discrete sum as a continuous integral corresponds to substituting the DTFT with a continuous Fourier transform. Replacing the periodic output of the DTFT with the aperiodic continuous FT, requires that the periodic convolution from \eqref{eq:dtft_conv_int} must be approximated as aperiodic by updating the integral limits to $\pm \infty$. 
The aperiodic convolution approximates the periodic one well, provided that the approximation of the SWB kernel \eqref{eq:rd_ap_int} decays to a negligible value before the first alias position. The positions of the aliases are given by the periodicity of the convolution, which is $1 / d_{\mathrm{s}_{\lambda}}$. 
This condition can be written as $R_{(D + d_{\mathrm{s}})}(\Delta u, \Delta u - u_{\tau}) \approx 0$ for $|u_{\tau}| > 1 / d_{\mathrm{s}_{\lambda}}$.
However, this stringent condition can be relaxed by observing that the  $\mathrm{AF_{NB}(u_{\tau})}$ in \eqref{eq:dtft_conv_int_lim} acts as a spatial filter which modulates the SWB kernel \eqref{eq:rd_int_approx}. Therefore, the SNB AF reduces the approximation error due to aliasing by acting as a spatial anti-aliasing filter. Therefore, the accuracy of the total SWB AF approximation depends on the mainlobe width and sidelobe levels of the $\mathrm{AF_{SNB}}$ at the alias positions given by $u_{\tau} = n / d_{\mathrm{s}_{\lambda}}$, where $n = \pm 1, \pm 2, \ldots$.
For a linear aperture of length $D_{\lambda}$ with uniform spatial weighting the mainlobe of $\mathrm{AF_{SNB}}(u_{\tau})$ is approximately to $1/ D_{\lambda}$. The aliases of  $R_{M}$ occur at $u_{\tau} = \Delta u + n/d_{\mathrm{s}_{\lambda}}$ for $n = \pm1, \pm2, \ldots$. For the alias contribution to be negligible, the mainlobe of the $\mathrm{AF_{SNB}}$ must be much narrower than the alias spacing $1 / D_{\lambda} \ll 1 / d_{\mathrm{s}_{\lambda}}$. 
Since $D_{\lambda} = (M-1) d_{\mathrm{s}_{\lambda}}$ the criterion can be simplified to $M \gg 1$. Therefore, the SWB AF approximation is valid as long as the number of antennas $M$ is sufficiently large. When considering windowing in the spatial domain, a window will effectively broaden the $\mathrm{AF_{SNB}}$ mainlobe while suppressing sidelobes. Depending on the applied window, mainlobe broadening must be accounted for and can lead to a scaling of the criterion by a factor of $N$-fold. For example, the Hamming and Hann windows broaden the mainlobe by a factor of 2 and the updated criterion can be written as $M \gg 2$, which is still negligible given the considered large apertures.
The smoothness criterion of $R_{M}$ introduced in \eqref{eq:rd_int_approx} guarantees accuracy of the SWB kernel approximation. However, even when this criterion is violated, the $M \gg 1$ criterion ensures that the alias errors in $R_D$ remain negligible in the convolution output via the spatial filtering by the SNB AF. 
This makes the approximation of the SWB AF robust and applicable even for large antenna spacing and fractional bandwidths as long as the number of antennas $M$ is sufficiently large.
Given the approximation, the SWB AF from \eqref{eq:dtft_conv_int} can be written as an aperiodic convolution as follows
\begin{align}
    \label{eq:infty_conv}
    \mathrm{AF_{SWB}}(\Delta u) 
    & \overset{M \gg 1}{\approx} \int_{-\infty}^{\infty} 
    W\left( v \right)
    R_{(D + d_{\mathrm{s}})}\left( \Delta u, -\frac{d_{\mathrm{s}}}{\lambda_{\mathrm{c}}} \Delta u - v \right)\, dv.
\end{align}
As in \eqref{eq:dtft_conv_int_lim}, expressing the convolution from \eqref{eq:infty_conv} in the directional sine domain $u_{\tau}$ via substitution $v = - \frac{d_{\mathrm{s}} }{\lambda_{\mathrm{c}}}u_{\tau}$ yields
\begin{align}
    \label{eq:wb_af_int_u_domain}
    &\mathrm{AF_{SWB}}(\Delta u) 
    \overset{M \gg 1}{\approx} \int_{-\infty}^{\infty} 
    \mathrm{AF_{SNB}}(u_{\tau}) \\
    & \qquad \qquad \qquad \qquad \times \tilde{R}_{(D + d_{\mathrm{s}})} \left( \Delta u, -\frac{d_{\mathrm{s}}}{\lambda_{\mathrm{c}}} \left( \Delta u - u_{\tau} \right) \right)\, du_{\tau}, \nonumber 
\end{align}
where $\tilde{R}_{(D + d_{\mathrm{s}})}(\Delta u, v) = \frac{d_{\mathrm{s}}}{\lambda_{\mathrm{c}}}  R_{(D + d_{\mathrm{s}})}(\Delta u, v)$ denotes the scaled SWB kernel. 
Evaluating the scaled SWB kernel \eqref{eq:rd_ap_int} for argument $v = - \frac{d_{\mathrm{s}}}{\lambda_{\mathrm{c}}} \left(\Delta u - u_{\tau} \right)$ gives
\begin{align}
    & \tilde{R}_{(D + d_{\mathrm{s}})}\left( \Delta u, -\frac{d_{\mathrm{s}}}{\lambda_{\mathrm{c}}} \left(\Delta u - u_{\tau} \right) \right) = \\
    &\qquad = \frac{1}{\lambda_{\mathrm{c}}} \int_{-(D + d_{\mathrm{s}}) / 2}^{(D + d_{\mathrm{s}}) / 2}  R \left( \frac{ x \Delta u }{c} \right) e^{j 2\pi \frac{x}{\lambda_{\mathrm{c}}} \left( \Delta u - u_{\tau} \right)} \, dx \nonumber.
\end{align}
Finally, by normalizing the integration variable, which is position on the aperture by wavelength via the substitution $x = x_{\lambda} \lambda_{\mathrm{c}}$, the $1/\lambda_{\mathrm{c}}$ scaling is absorbed as follows
\begin{align}
    \label{eq:r_d_norm}
    &\tilde{R}_{(D_{\lambda} + d_{\mathrm{s}_{\lambda}})} \left( \Delta u, -\frac{d_{\mathrm{s}}}{\lambda_{\mathrm{c}}} \left( \Delta u - u_{\tau} \right) \right) = \\
    &\qquad  = \int_{-(D_{\lambda} + d_{\mathrm{s}_{\lambda}}) / 2}^{(D_{\lambda} + d_{\mathrm{s}_{\lambda}}) / 2}  R \left( \frac{ x_{\lambda} \Delta u }{f_{\mathrm{c}}} \right) e^{j 2\pi x_{\lambda} \left( \Delta u - u_{\tau} \right)} \, dx_{\lambda}. \nonumber
\end{align}

\subsection{Continuous spatially wideband kernel with windowing in frequency domain}
Next, to provide a generic model for an arbitrary amplitude spectrum consider a uniform amplitude spectrum $|S(f)| = \frac{1}{\sqrt{B}}$ windowed by a function $W(f)$ with normalized amplitude as follows $\int_{-B/2}^{B/2} |W(f)|\, df = 1$. In the presence of windowing, the signal autocorrelation function from \eqref{eq:autocorr} becomes 
\begin{align}
    \label{eq:autocorr_wdw}
    R\left( \tau \right) 
    &=\frac{1}{B} \int_{-B/2}^{B/2}
    W(f) e^{j2\pi f \tau } \, df.
\end{align}
The continuous windowing function can be represented by it Fourier transform pair as
\begin{align}
    \label{eq:wdw_ft}
    W(f) = \int_{-\infty}^{\infty} w(t) e^{-j 2 \pi f t} \, dt.
\end{align}
Expanding the windowed signal autocorrelation from \eqref{eq:autocorr_wdw} using the FT of the window from \eqref{eq:wdw_ft} yields
\begin{align}
    \label{eq:wdw_r_tau_sinc}
    R\left( \tau \right) 
    &=\frac{1}{B} \int_{-B/2}^{B/2}
    \left( \int_{-\infty}^{\infty} w(t) e^{-j 2 \pi f t} \, dt \right) e^{j2\pi f \tau } \, df \\
    &= \int_{-\infty}^{\infty} w(t) \sinc{ \left( B \left( \tau - t \right) \right) 
    }\, dt \nonumber. 
\end{align}
Substituting the expanded windowed autocorrelation function from \eqref{eq:wdw_r_tau_sinc} into the expression of the SWB kernel \eqref{eq:r_d_norm} gives
\begin{align}
    \label{eq:rd_norm_wdw}
    &\tilde{R}_{(D_{\lambda} + d_{\mathrm{s}_{\lambda}})}\left( \Delta u, -\frac{d_{\mathrm{s}}}{\lambda_{\mathrm{c}}} \left( \Delta u - u_{\tau} \right) \right) = \\
    &= \int_{-\infty}^{\infty} w(t) \Bigg[ \int_{-(D_{\lambda} + d_{\mathrm{s}_{\lambda}}) / 2}^{(D_{\lambda} + d_{\mathrm{s}_{\lambda}}) / 2} 
    \sinc{ \left( B \left( \frac{ x_{\lambda} \Delta u }{f_{\mathrm{c}}} - t \right) \right) 
    } \nonumber \\
    & \qquad \qquad \qquad \quad \times  e^{j 2\pi x_{\lambda} \left( \Delta u - u_{\tau} \right)} \, dx_{\lambda} \Bigg] \, dt . \nonumber
\end{align}
The derivation of the SWB kernel in \eqref{eq:rd_norm_wdw} can be easily extended to a discretized spectrum, for example, OFDM of bandwidth $B$ with $K$ subcarriers symmetrically distributed around 0.
For a continuous spectrum, it is possible to obtain a closed-form expression of the aperture integral by expressing it in terms of sine and cosine integral functions. However, for clarity, the detailed derivation for an arbitrary window is omitted. Instead, numerical results based on closed-form expressions are provided to offer insight into the effects of windowing on the SWB kernel shape. 
For uniform window $W(f) = 1/B$, $w(t) = \sinc{(Bt)}$ the SWB kernel from \eqref{eq:rd_norm_wdw} simplifies to
\begin{align}
    \label{eq:rd_norm_uni_spec}
    &\tilde{R}_{(D_{\lambda} + d_{\mathrm{s}_{\lambda}})} \left( \Delta u, -\frac{d_{\mathrm{s}}}{\lambda_{\mathrm{c}}} \left( \Delta u - u_{\tau} \right) \right) =  \\
    &= \frac{1}{B_{\mathrm{f}} \Delta u}
    \Bigg( 
    \Si{\left( (D_{\lambda} + d_{\mathrm{s}_{\lambda}}) \left( \frac{B_{\mathrm{f}}}{2}\Delta u + \Delta u - u_{\tau} \right) \right)} \nonumber  \\
    & \qquad \quad + \Si{\left( (D_{\lambda} + d_{\mathrm{s}_{\lambda}}) \left( \frac{B_{\mathrm{f}}}{2}\Delta u - \Delta u + u_{\tau} \right) \right)}
    \Bigg) \nonumber .
\end{align}

\begin{figure}[tb]
    \centering
    \includegraphics[width=\linewidth]{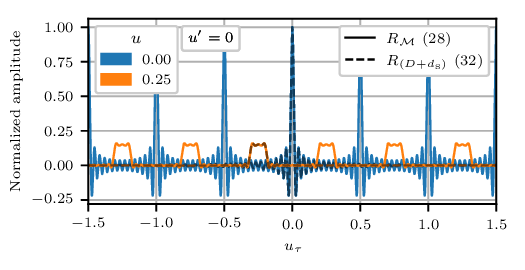}
    \caption{Discrete vs continuous spatially wideband kernel for uniform spectrum, $u'=0$, $D_{\lambda}=52$, $d_{s_{\lambda}} = 2$ and $B_{\mathrm{f}} = 0.5$.}
    \label{fig:rd_vs_approx}
\end{figure}

Fig. \ref{fig:rd_vs_approx} shows the comparison between the discrete SWB kernel and its continuous approximation for a uniform spectrum. As discussed before, the continuous approximation is aperiodic, while the discrete kernel exhibits aliases.
For large values of $B_{\mathrm{f}}$ and $\Delta u$, it is possible that the aliases will overlap and affect the accuracy of the kernel approximation.
\begin{figure}[tb]
    \centering
    \includegraphics[width=\linewidth]{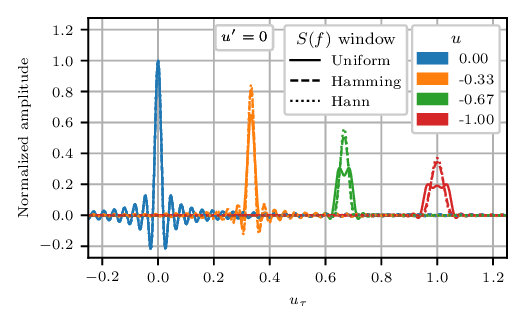}
    \caption{Spatially wideband kernel $\tilde{R}_{(D_{\lambda} + d_{\mathrm{s}_{\lambda}})}$ \eqref{eq:rd_norm_wdw} as a function of the convolution variable $u_\tau$ for different spectrum windows, $u'=0$, $D_{\lambda} = 52$ and $B_{\mathrm{f}} = 0.1$.}
    \label{fig:rd_vs_u_wdw}
\end{figure}

Fig. \ref{fig:rd_vs_u_wdw} illustrates the SWB kernel from \eqref{eq:rd_norm_wdw} for varying $u$ and different spectrum windows. The plot is limited to a few $u$ values as $\tilde{R}$ is an even function of $u$ centered around $u'$.
As $|u|$ increases, the SWB kernel widens and its amplitude decreases. Applying a window in the spectral domain reduces the spread of the kernel and suppresses its sidelobes, making it more concentrated around $u$. From the perspective of the SWB kernel, spectral windowing reduces the mainlobe width and sidelobes of the kernel, thereby relaxing the anti-aliasing constraint from the aperiodic convolution approximation of the SWB AF \eqref{eq:infty_conv}.
Therefore, the uniform spectrum window can be considered as the worst-case scenario of the SWB AF approximation, introducing the most severe aliasing in the SWB kernel approximation.
The widening of $\tilde{R}_{(D_{\lambda} + d_{\mathrm{s}_{\lambda}})}$ can be considered beneficial in terms of sidelobe and grating-lobe attenuation. A wider convolution kernel reduces the amplitude of sidelobes and grating lobes by averaging them over a larger $u_{\tau}$ span.
Therefore, spectrum windowing degrades sidelobe and grating-lobe attenuation because it results in a narrower, more concentrated SWB kernel. Intuitively, spectrum windowing applies a weighting across the bandwidth, attenuating the contribution of edge frequencies in the SWB AF integral from \eqref{eq:wb_af_sep}.

\section{Uniform sparse arrays}
The SWB AF approximation introduced in Sec. \ref{sec:gen_wb_af_approx} is derived for uniform linear antenna arrays with an arbitrary element spacing. In this section, the accuracy of the SWB AF approximation is evaluated across element spacing and fractional bandwidths.
Moreover, bandwidth-enabled grating lobe suppression is investigated for uniform sparse arrays with element spacing greater than $\lambda_{\mathrm{c}}/2$.

\begin{figure}[tb]
    \centering
    \includegraphics[width=\linewidth]{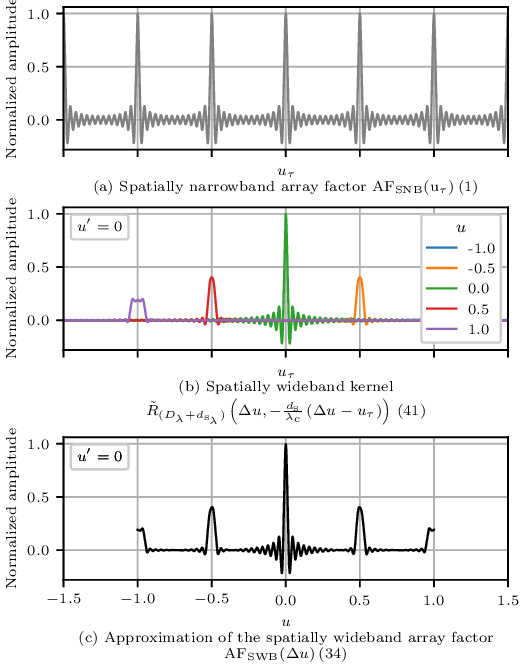}
    \caption{Illustration of spatially variant convolution and its components yielding the approximation of the spatially wideband AF for a uniform spectrum window, $u'=0$, $D_{\lambda}=52$, $d_{s_{\lambda}} = 2$ and $B_{\mathrm{f}} = 0.1$.}
    \label{fig:conv_visualization}
\end{figure}

Fig. \ref{fig:conv_visualization} illustrates the approximation of the SWB AF and its individual components when expressed as aperiodic spatially variant convolution from \eqref{eq:wb_af_int_u_domain}.
To illustrate the averaging effect of bandwidth on the SWB AF, the spacing between antenna elements is sparse $d_{\mathrm{s}_{\lambda}} = 2$, which introduces grating lobes with periodicity $1 / d_{\mathrm{s}_{\lambda}} = 0.5$ in the $u$ domain.
Fig. \ref{fig:conv_visualization}a shows the SNB AF 
\eqref{eq:nb_af}, with periodic grating lobes of equal amplitude due to sparse antenna spacing.
Fig. \ref{fig:conv_visualization}b illustrates the SWB kernel as a function of $u_\tau$ for a few selected $u$ values.
Finally, Fig. \ref{fig:conv_visualization}c shows the resulting SWB AF, which is obtained by multiplying the SNB AF in Fig. \ref{fig:conv_visualization}a with the SWB kernel from Fig. \ref{fig:conv_visualization}b and integrating the result over $u_\tau$ for each $u$ value, as per \eqref{eq:wb_af_int_u_domain}. The modulation of the SWB kernel by a SNB AF allows for the accurate estimation of the SWB AF even in the presence of aliasing errors arising from the approximation in \eqref{eq:rd_int_approx}.
Notice that due to the widening and reduced amplitude of the SWB kernel as $|u|$ increases, the grating lobes in the SWB AF are also widened and their peak amplitudes are reduced.

\begin{figure}[b!]
    \centering
    \includegraphics[width=\linewidth]{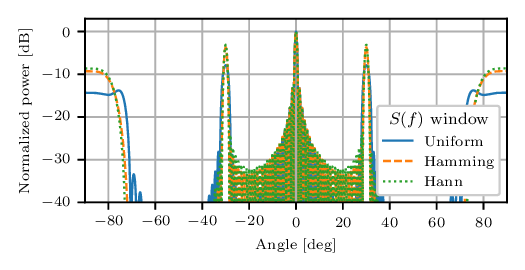}
    \caption{Spatially wideband antenna AF \eqref{eq:wb_af_r_sum} for different spectrum windows for $u'=0$, $D_{\lambda}=52$, $d_{s_{\lambda}} = 2$ and $B_{\mathrm{f}} = 0.1$.}
    \label{fig:wb_af_vs_wdw}
\end{figure}

Fig. \ref{fig:wb_af_vs_wdw} shows the SWB AF \eqref{eq:wb_af_r_sum} for a few selected spectrum windows and the same system parameters as in Fig. \ref{fig:conv_visualization}. 
By observing the grating lobes power and shape, it becomes apparent that windowing degrades the bandwidth ability to spread the power of the sidelobes, as it results in a more concentrated SWB kernel from \eqref{eq:rd_norm_wdw}, as shown in Fig. \ref{fig:rd_vs_u_wdw}.
Therefore, a uniform spectrum window offers the best-case scenario when considering bandwidth as a means of suppressing the grating lobes.

\begin{figure}[tb]
    \centering
    \includegraphics[width=\linewidth]{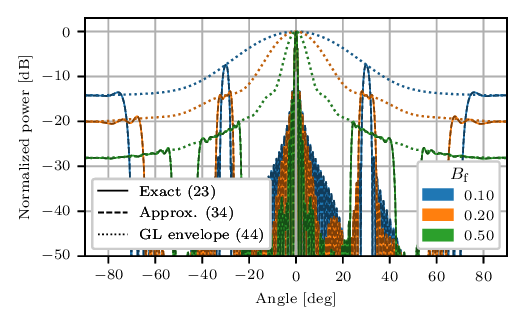}
    \caption{Comparison between the exact SWB AF and the approximation for a uniform spectrum window, $u'=0$, $D_{\lambda}=52$, $d_{s_{\lambda}} = 2$ and selected fractional bandwidths.}
    \label{fig:wb_af_vs_apx_vs_fbw}
\end{figure}

\begin{figure}[tb]
    \centering
    \includegraphics[width=\linewidth]{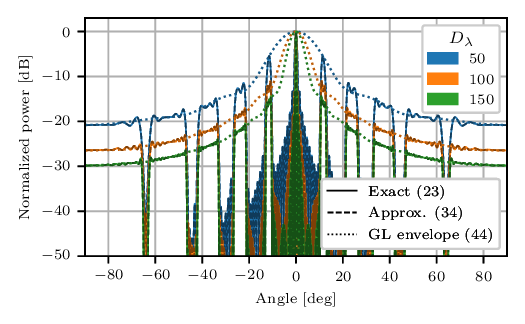}
    \caption{Comparison between the exact SWB AF and the approximation across different aperture sizes for a uniform spectrum window, $u'=0$, $B_{\mathrm{f}} = 0.2$, $d_{s_{\lambda}} = 5$.}
    \label{fig:wb_af_vs_apx_vs_dap}
\end{figure}

Fig. \ref{fig:wb_af_vs_apx_vs_fbw} compares the exact SWB AF with the approximation for different fractional bandwidths and a uniform spectrum window. The larger values of $B_{\mathrm{f}}$ allow for better suppression of the grating lobes and sidelobes at the cost of broadening their angular spread. Fig. \ref{fig:wb_af_vs_apx_vs_dap} illustrates the accuracy of the SWB AF approximation for different aperture sizes and larger element spacing. 
The SWB AF approximation remains accurate even for large values of fractional bandwidth $B_{\mathrm{f}}$ and $d_{\mathrm{s}_{\lambda}}$. For large products of $B_{\mathrm{f}} d_{\mathrm{s}_{\lambda}}$, the approximation of the SWB kernel might be inaccurate due to aliasing, as discussed in Fig. \ref{fig:rd_vs_approx}. However, in the total SWB AF, the error due to aliasing is compensated by the SNB AF that acts as the spatial filter in the convolution, reducing the error originating from the aliases.

\subsection{Approximation of grating lobes envelope}
Since the uniform amplitude spectrum (window) offers the best suppression of the grating lobes, in the following introduces another approximation of the SWB kernel for a uniform spectrum. Next, based on the approximation, the closed-form expression of the grating lobe (GL) amplitude envelope is obtained as a function of their position and bandwidth-aperture product.
Consider that the  argument of the $\Si{}$ function in \eqref{eq:rd_norm_uni_spec} is sufficiently large $\left( (D_{\lambda} + d_{\mathrm{s}_{\lambda}}) \left( \frac{B}{2 f_{\mathrm{c}}}\Delta u \pm \left( \Delta u - u_{\tau} \right) \right) \right) \gg 1$ so that the $\Si{}$ function can be approximated by the sign function as follows $\Si(x) \approx \frac{1}{2} \sgn(x)$. This is a valid assumption provided that $D_{\lambda} B_{\mathrm{f}} |\Delta u| \gg 1$.
Approximating the sine integral as a sign function enables approximating the SWB kernel from \eqref{eq:rd_norm_uni_spec} as a rectangular function as follows

\begin{align}
    \label{eq:rd_norm_rect}
    &\tilde{R}_{\rect{}}\left( \Delta u, -\frac{d_{\mathrm{s}}}{\lambda_{\mathrm{c}}} \left( \Delta u - u_{\tau} \right) \right)
    \overset{D_{\lambda} B_{\mathrm{f}} |\Delta u| \gg 0}{\approx} \nonumber \\
    &\qquad \qquad \qquad \overset{D_{\lambda} B_{\mathrm{f}} |\Delta u| \gg 0}{\approx}  \frac{f_{\mathrm{c}}}{B |\Delta u|} \rect{\left( \frac{\Delta u - u_{\tau}}{ \frac{B}{f_{\mathrm{c}}} |\Delta u| }\right)}.
\end{align}
The SWB kernel is approximated as a rectangular function of width $B_{\mathrm{f}}|\Delta u|$, which increases with both the bandwidth and angle. Note that the total energy of the kernel is conserved and equal to unity. As the kernel widens, the amplitude decreases accordingly. Therefore, the reduction in the amplitude of the grating lobes is achieved by broadening their main lobes and spreading the energy over a wider angular range, as shown in Fig. \ref{fig:wb_af_vs_apx_vs_fbw}.

To approximate how the SWB kernel reduces the grating lobe power, consider a single isolated grating lobe. As the grating lobes are the spatial aliases of the mainlobe, the AF of the grating lobe can be approximated by that of the equivalent array with the same total aperture and spacing that satisfies the Nyquist criterion $d_{\mathrm{s}_{\lambda}} \leq 0.5$ 
\begin{align}
    \label{eq:af_grating_lobe}
    \mathrm{AF_{SNB}}(u_{\tau})
    &=\frac{\sin{\left( \pi M d_{\mathrm{s}_{\lambda}} u_{\tau} \right)}}
    {M\sin{\left( \pi d_{\mathrm{s}_{\lambda}} u_{\tau} \right)}} 
    \overset{\substack{M\gg 1 \\ d_{\mathrm{s}_{\lambda}} \leq 0.5}}{\approx} \sinc{\left( Md_{\mathrm{s}_{\lambda}} u_{\tau} \right)}.
\end{align}
Considering a single isolated grating lobe simplifies the analysis, enabling to obtain a closed-form expression of the grating lobe amplitude as a function of the angle and bandwidth-aperture product. This approach is based on the assumption that the contributions from the adjacent grating lobes and their sidelobes are negligible. That is, the width of the SWB kernel from \eqref{eq:rd_norm_rect} is much smaller than the spacing between the grating lobes $\frac{B_{\mathrm{f}}}{2} |\Delta u| \ll \frac{1}{d_{\mathrm{s}_{\lambda}}}$. 
The amplitude of the grating lobe located at angle $u_g$ is obtained by substituting the approximation of the SWB kernel \eqref{eq:rd_norm_rect} and the grating lobe AF \eqref{eq:af_grating_lobe} into \eqref{eq:wb_af_int_u_domain} and evaluating it for the grating lobe position $\Delta u = u_g$
\begin{align}
    \label{eq:gl_env_approx}
    &\mathrm{AF_{SWB}}(u_g)  \overset{M \gg 1}{\approx} \nonumber \\
    & \qquad  \overset{M \gg 1}{\approx} \frac{f_{\mathrm{c}}}{B |u_g|} \int_{-\infty}^{\infty} 
    \sinc{\left( M d_{\mathrm{s}_{\lambda}} u_{\tau} \right)}
    \rect{\left( \frac{u_g - u_{\tau}}{ \frac{B}{f_{\mathrm{c}}} |u_g| }\right)} \, d u_{\tau}   \nonumber \\
    &\qquad = \frac{2}{B_{\mathrm{f}} (D_{\lambda} + d_{\mathrm{s}_{\lambda}}) |u_g|} 
    \Si{\left( (D_{\lambda} + d_{\mathrm{s}_{\lambda}}) \frac{B_{\mathrm{f}}}{2} |u_g| \right)}. 
\end{align}
\begin{figure}[t]
    \centering
    \includegraphics[width=\linewidth]{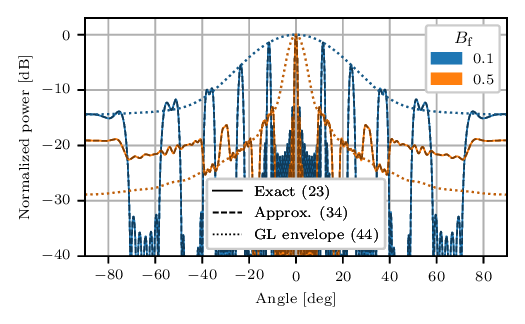}
    \caption{Comparison between the exact and approximated SWB AF for a large antenna spacing and fractional bandwidth uniform spectrum window, $u'=0$, $D_{\lambda} = 50$, $d_{s_{\lambda}} = 5$.}
    \label{fig:gl_env_vs_fbw_inacc}
\end{figure}
The approximation of the grating lobes amplitude can be considered accurate for $u_g$ in the following range $\frac{1}{D_{\lambda} B_{\mathrm{f}}} \ll |u_g| \ll \frac{2} {d_{\mathrm{s}_{\lambda}} B_{\mathrm{f}}}$ where the lower bound is determined by \eqref{eq:rd_norm_rect}.
Fig. \ref{fig:wb_af_vs_apx_vs_fbw} and Fig. \ref{fig:wb_af_vs_apx_vs_dap} both include the approximation of the envelope of the grating lobe amplitudes. The approximation remains accurate for large fractional bandwidths and apertures, given that the bandwidth-spacing $B_{\mathrm{f}} d_{\mathrm{s}_{\lambda}}$ product is sufficiently small. If the approximation criterion is not satisfied, the wide SWB kernel captures contributions from other grating lobes, resulting in increased sidelobe levels. Fig. \ref{fig:gl_env_vs_fbw_inacc} illustrates the scenario when the grating lobe amplitude approximation fails for a large $B_{\mathrm{f}} d_{\mathrm{s}_{\lambda}} = 2.5$. The angular region where the approximation holds is $|u_g| \ll 0.8 $, which corresponds to angle $\theta_g \ll 53 \degree$. For large $B_{\mathrm{f}} d_{\mathrm{s}_{\lambda}}$ products, the $|u_g|$ region where the approximation holds is greatly reduced.
Note that the SWB AF approximation from \eqref{eq:wb_af_int_u_domain} remains accurate. However, the grating lobe suppression is no longer as effective, as the width of the SWB kernel captures contributions from multiple grating lobes, resulting in degraded suppression (averaging). Moreover, the spreading of the grating lobes by the wide SWB kernel results in an increased sidelobe floor, removing the notches and nulls in the SWB AF. Increasing bandwidth in systems with large spacing allows to suppress the grating lobe levels at the cost of spreading their energy and creating a sidelobe floor.

\begin{figure}[h]
    \centering
    \includegraphics[width=\linewidth]{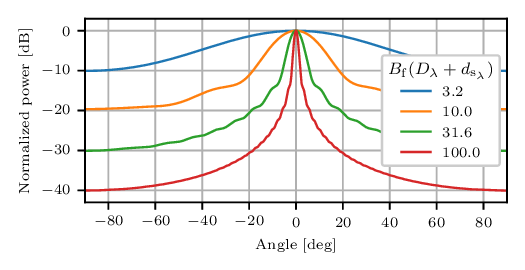}
    \caption{Grating lobe amplitude envelope as a function of angle for $u'=0$, $\theta' = 0 \degree$, uniform spectrum and different bandwidth-aperture products.}
    \label{fig:gl_env_vs_bf_dap_prod}
\end{figure}

Fig. \ref{fig:gl_env_vs_bf_dap_prod} shows the envelope of the grating lobes for different values of the bandwidth-aperture product. This plot illustrates the achievable levels of grating lobe suppression, provided that the approximation conditions from \eqref{eq:gl_env_approx} are satisfied.

\section{Conclusion}
This work extends the AF concept to the SWB regime. For fully populated arrays with element spacing that satisfies the Nyquist criterion, closed-form expressions are derived. These expressions quantify how bandwidth reduces the sidelobe level compared to the SNB AF, including the effect of spectral and spatial windowing. A universal approximation of the SWB AF is introduced by expressing the SWB AF as a spatially variant convolution. This approximation is suitable for a wide range of antenna element spacings and bandwidths. It offers insight into how the SWB kernel spreads and averages the sidelobes and grating lobes. Finally, the approximation of the envelope of the grating lobes is introduced, enabling prediction of the grating lobe levels in SWB AF as a function of their angular position and the bandwidth-aperture product.

\bibliographystyle{IEEEtran}
\bibliography{IEEEabrv.bib, biblio.bib}

\end{document}